%% file: main_ACL.tex
\pdfoutput=1

\documentclass[11pt]{article}

\usepackage[final]{acl2023}
\usepackage{array}
\usepackage{multirow}
\usepackage{graphicx}
\usepackage{comment}
\usepackage{booktabs}
\usepackage{times}
\usepackage{latexsym}
\usepackage{enumitem}

\usepackage[T1]{fontenc}

\usepackage[utf8]{inputenc}

\usepackage{microtype}

\usepackage{inconsolata}

\title{Disentangling Perceptions of Offensiveness: Cultural and Moral Correlates}

\author{Aida Davani \\
  Google Research \\
  \texttt{aidamd@google.com} \\\And
  Mark Díaz \\
  Google Research\\
  \texttt{markdiaz@google.com} \\\And
  Dylan Baker\\
  Distributed AI Research Institute\\
  \texttt{dylan@dair-institute.org } \\\AND
  Vinodkumar Prabhakaran\\
  Google Research \\
  \texttt{vinodkpg@google.com}\\
  }

\begin{document}
\maketitle
\begin{abstract}
\input{sections/abstract}

\end{abstract}

\input{sections/intro}

\section*{Data Collection}
\input{sections/data}

\section*{Study 1: Geo-cultural Differences in Offensiveness}
\input{sections/study1}

\section*{Study 2: Moral Foundations of Offensiveness}

\input{sections/study2}

\section*{Study 3: Implications for Responsible AI}

\input{sections/study3}

\section*{General Discussion}
\input{sections/discussion}

\bibliographystyle{acl_natbib}
\bibliography{acl-bib}

\appendix

\section{Appendix}
\label{sec:appendix}
\input{sections/appendix}

\end{document}

%% file: sections/abstract.tex

Perception of offensiveness is inherently subjective, shaped 
by the lived experiences and socio-cultural values of the perceivers.
Recent years have seen substantial efforts to build AI-based tools that can detect offensive language at scale, as a means to moderate social media platforms, and to ensure safety of conversational AI technologies such as ChatGPT and Bard.
However, existing approaches 
treat this
task as a technical endeavor, built on top of data annotated for offensiveness by a global crowd workforce without any attention to the crowd workers' provenance or the values their perceptions reflect. 
We argue that cultural and psychological factors play a vital role in the cognitive processing of offensiveness,
which is critical to consider
in this context.
We re-frame the task of determining offensiveness as essentially a matter of moral judgment --- deciding the boundaries of ethically wrong vs. right language
within an implied set of socio-cultural norms.
Through a large-scale cross-cultural study based on 4309 participants from 21 countries across 8 cultural regions, we demonstrate substantial cross-cultural differences in perceptions of offensiveness. 
More importantly, we find that individual moral values
play a crucial role in shaping these variations: moral concerns about Care and Purity are significant mediating factors driving cross-cultural differences.
These insights are of crucial importance as we build AI models for the pluralistic world, where the values they espouse should aim to respect and account for moral values in diverse geo-cultural contexts.

%% file: sections/intro.tex
As conversational Artificial Intelligence (AI) technologies 
are becoming ubiquitous,\footnote{such as ChatGPT (\url{chat.openai.com}) and Bard (\url{bard.google.com})} 
there is an increasing urgency to equip them with safety guardrails that prevent inadvertent generation of offensive and hateful content, as evident in many recent academic and governmental calls for action \cite{bai2022training,glaese2022improving,european2020digital,wh2023}. Technologies developed for supporting such guardrails essentially rely on human assessments of language \cite{schmidt2017survey}
that are emulated through
reinforcement learning \cite{ouyang2022training} or supervised modeling \cite{garg2023handling}. The objective of this paper is to demonstrate that such modeling approaches should consider the socio-cultural and moral aspects of language interpretation that impact annotators' subjective judgments, but are not captured through existing data annotation practices.

\begin{figure}[]
    \centering
    \includegraphics[width=.9\linewidth]{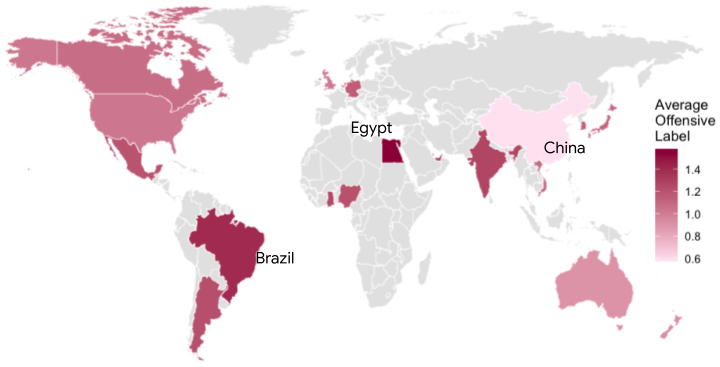}
    \caption{\textbf{Distribution of labels provided from different countries.} Annotators from China, Brazil, and Egypt provided significantly different labels.}
    \label{fig:countries}
\end{figure}

Designing Natural Language Processing (NLP) tools for detecting offensive or toxic text has long been an active area of research, originally aimed at automating online content moderation at scale \cite{wulczyn2017ex,founta2018large}. However, traditional NLP approaches largely overlooked cultural and individual factors that shape annotators' perspectives on what is deemed offensive \cite{aroyo2015truth,talat2016you, salminen2019online,uma2021learning,prabhakaran2021releasing}.
This oversight is especially troubling when these tools are developed as safety guardrails for conversational AI, considering the pace and scale at which these technologies are being adopted across geo-cultural contexts. 
The risk of cultural skews becomes particularly salient in this context, considering that the crowd-sourced workforce, essential for annotation efforts, is disproportionately made up of workers from the Global South \cite{geva2019modeling,aroyo2023reasonable}.
Tailoring safety guardrails to only some subgroups' values and preferences would in effect marginalize perspectives and concerns of others \cite{salminen2019online}.


Furthermore, disagreements on what is offensive expand beyond mere differences in socio-cultural backgrounds. For instance, 
the intricate interplay of social media content moderation and principles of freedom of speech brings the task of offensive language detection into the realm of moral and political deliberation (instances of such discussions can be found in \cite{balkin2017digital,brannon2019free,kiritchenko2021confronting}). 
More generally, individuals might systematically disagree on notions of offensiveness, reflecting the complexity of beliefs and values that shape their perspectives and judgments within any given cultural context.
Therefore, we hypothesize that the high divergence in annotators' perceptions of offensiveness \cite{prabhakaran2021releasing} can be traced back to individuals' diverse values besides the cultural and social norms that dictate the boundaries of acceptable language within a society.

In particular, we argue that these values are shaped by what is considered as morally right or wrong within a culture or by individuals.
Annotators from diverse cultural backgrounds
may
perceive offensiveness in distinct ways and this variation in cultural perspectives can be attributed to both the social norms prevalent in their societies,
and the moral values held by individuals within the given socio-cultural context. To capture variances in moral values we rely on the Moral Foundation Theory \cite{graham2013moral} which proposes six innate and universally available \textit{foundations} that form notions of moral reasoning. As per this theory, moral foundations shape social behaviors and judgments about what is considered as morally right or wrong across different cultures and individuals.

Towards this goal, we conduct a two-pronged study based on a large-scale cross-cultural language annotation experiment, where we asked participants from eight different cultural regions, balanced across various socio-demographic groups, to annotate offensiveness in social media posts curated from a popular NLP dataset. In study 1, we demonstrate that perceptions of offensiveness vary significantly across geo-cultural regions, even after controlling for gender, age, and socio-economic status of the participants, or whether or not they were given a specific definition for offensiveness. 
In order to investigate how individual moral values shape these differences in perceptions, we also had each of our participants fill out the Moral Foundations Questionnaire \cite{atari2022morality}. In study 2, we find that the cross-cultural differences in perceptions of offensiveness are significantly mediated by annotators' individual moral concerns, in particular, of Care and Purity, that vary across cultures.

We further demonstrate the real-world implications of these finding:
popular AI-based models that
analyze toxic language do preferentially align with perspectives of individuals associated with specific geo-cultural regions and moral values.
%
These findings emphasize the importance of understanding the cultural and psychological factors for evaluating safety in AI models and content moderation, highlighting the possible need for culturally-informed data collection and model training efforts. Our research further advocates for AI model alignment efforts that are informed by variations of moral values across cultures and individuals, suggesting a meaningful paradigm for value alignment beyond socio-demographic categorizations.

%% file: sections/data.tex
We recruited 4309 participants from 21 countries,\footnote{Argentina, Australia, Brazil, Canada, China, Egypt, Germany, Ghana, India, Japan, Mexico, Netherlands, New Zealand, Nigeria, Qatar, Singapore, South Korea, United Arab Emirates, United Kingdom, United States of America, Vietnam} across eight geo-cultural regions,\footnote{Arab Culture, Indian Cultural Sphere, Latin America, North America, Oceania, Sinosphere, Sub Saharan Africa, and Western Europe --- loosely based on the UN SDG groupings \url{https://unstats.un.org/sdgs/indicators/regional-groups} with minor modifications: combining Australia, NZ and Oceania to Oceania, and separating North America and Europe, to facilitate easier data collection.}
and asked them to (i) annotate offensiveness of a set of text items, and (ii) respond to a self-report measure of moral concerns. 
Half of the annotators were given a specific definition for ``extremely offensive language'' as \textit{profanity, strongly impolite, rude or vulgar language expressed with fighting or hurtful words in order to insult a targeted individual or group}, while the other half was not provided with a definition to create a control setting of participants who are expected to lean on their individual notion of offensiveness.

Recruitment criteria accounts for various demographic attributes: (1) \textit{Region of residence}: we recruited at least 450 participants from each of the eight regions,
(2) \textit{Gender}: within regions, we set a maximum limit of 60\% representation for each of the two binary genders,\footnote{Collecting non-binary gender information is not safe for annotators in many countries, so we limited this study to binary gender in our study to ensure consistency across countries.}
(3) \textit{Age}: in each region at most 60\% of participants are 18 to 30 years old and at least 15\% are 50 years old or older,
and (4) \textit{English fluency}: we only selected participants with high self-reported proficiency in reading and writing English.
Additionally, we collected participants' self-reported subjective socio-economic status \cite{adler2000relationship} as a potential confound.

We performed this study in the English language. We selected items from Jigsaw's Toxic Comments Classification dataset \cite{Jigsaw-toxic}, and the Unintended Bias in Toxicity Classification dataset \cite{Jigsaw-bias}, both of which consist of social media comments 
labeled for toxicity.
We built a dataset of $N_{items}$ = 4554 consisting of three categories of items sampled from the above datasets:
(1) a random set of items that are likely to be more ambiguous (50\%); 
(2) a balanced set of items that mention specific social group identities related to gender, sexual orientation, or religion (40\%); and
(3) a balanced set of items include different moral sentiments, identified through a moral language tagger trained on the MFTC dataset \cite{hoover2020moral}.
(more description in the Materials and Methods).

Each participant was tasked to label the offensiveness of 40 items (including 5 control items for quality check) on a 5-point Likert scale (from \textit{not offensive at all} to \textit{extremely offensive}). Those who failed the quality control check were removed, and not counted against our final set of 4295 participants (see Materials and Methods for more details). 
Each item in the final dataset had labels from at least three participants in each region who passed the control check. 


After annotation, participants were also asked to fill out the Moral Foundations Questionnaire \cite[MFQ-2;][]{graham2013moral,atari2022morality}, which assesses their moral values along six different dimensions. The questionnaire includes 36 statements to assess participants' priorities along six moral foundations that include their intuitions about: \textit{Care}: avoiding emotional and physical damage to another individual, \textit{Equality}: equal treatment and equal outcome for individuals, \textit{Proportionality}: individuals getting rewarded in proportion to their merit or contribution, \textit{Authority}: deference toward legitimate authorities and the defense of traditions, \textit{Loyalty}: cooperating with ingroups and competing with outgroups, and \textit{Purity}: avoiding bodily and spiritual contamination and degradation.
We aggregate each participant's responses to compute a value between 1 to 5 to capture their moral foundations along each of these dimensions.

%% file: sections/study1.tex

As Figure \ref{fig:countries} shows, the perceptions of offensiveness in our data varied significantly across participants from different geo-cultural regions. This trend was further confirmed by a one-way ANOVA test using a cross-classified mixed-level regression model with annotators as the first level, regions as the second level, and items as the crossed level that reported $F$(7,7515) = 31.47, $p$ < .001.

On average, participants from the Arab Culture ($M$ = 1.19, $SD$ = 1.48) and Latin America ($M$ = 1.13, $SD$ = 1.39) reported highest levels of offensive scores, while participants from Sinosphere ($M$ = 0.80, $SD$ = 1.22) and Oceania ($M$ = 0.80, $SD$ = 1.19) reported the least values of offensiveness. 

\begin{figure}[]
    \centering
    \includegraphics[width=\linewidth]{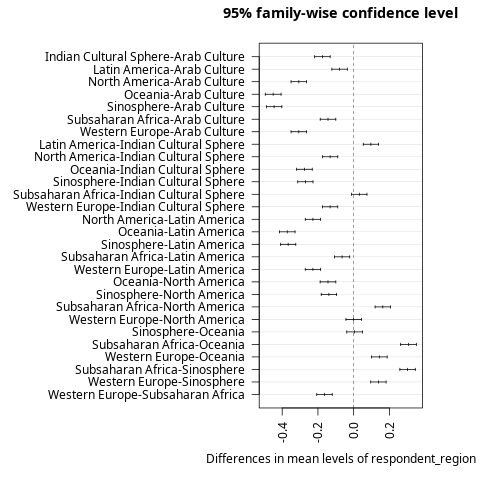}
    \caption{\textbf{Pairwise differences between geo-cultural regions in their perceptions of offensiveness.} 25 out of the 28 pairs of regions shows significant differences.}
    \label{fig_pairwise}
\end{figure}

Pairwise comparisons of regions show significant differences between 25 out of the 28 pairs of regions (See Figure~\ref{fig_pairwise}). In line with the above findings, annotations from Arab Culture and Latin America differed significantly from every other region.
For three pairs of regions (Sub-Saharan Africa and Indian Culture, Western Europe and North America, Sinosphere and Oceania) the annotations are not significantly different.


\begin{figure}[t]
    \centering
    \includegraphics[width=.47\textwidth]{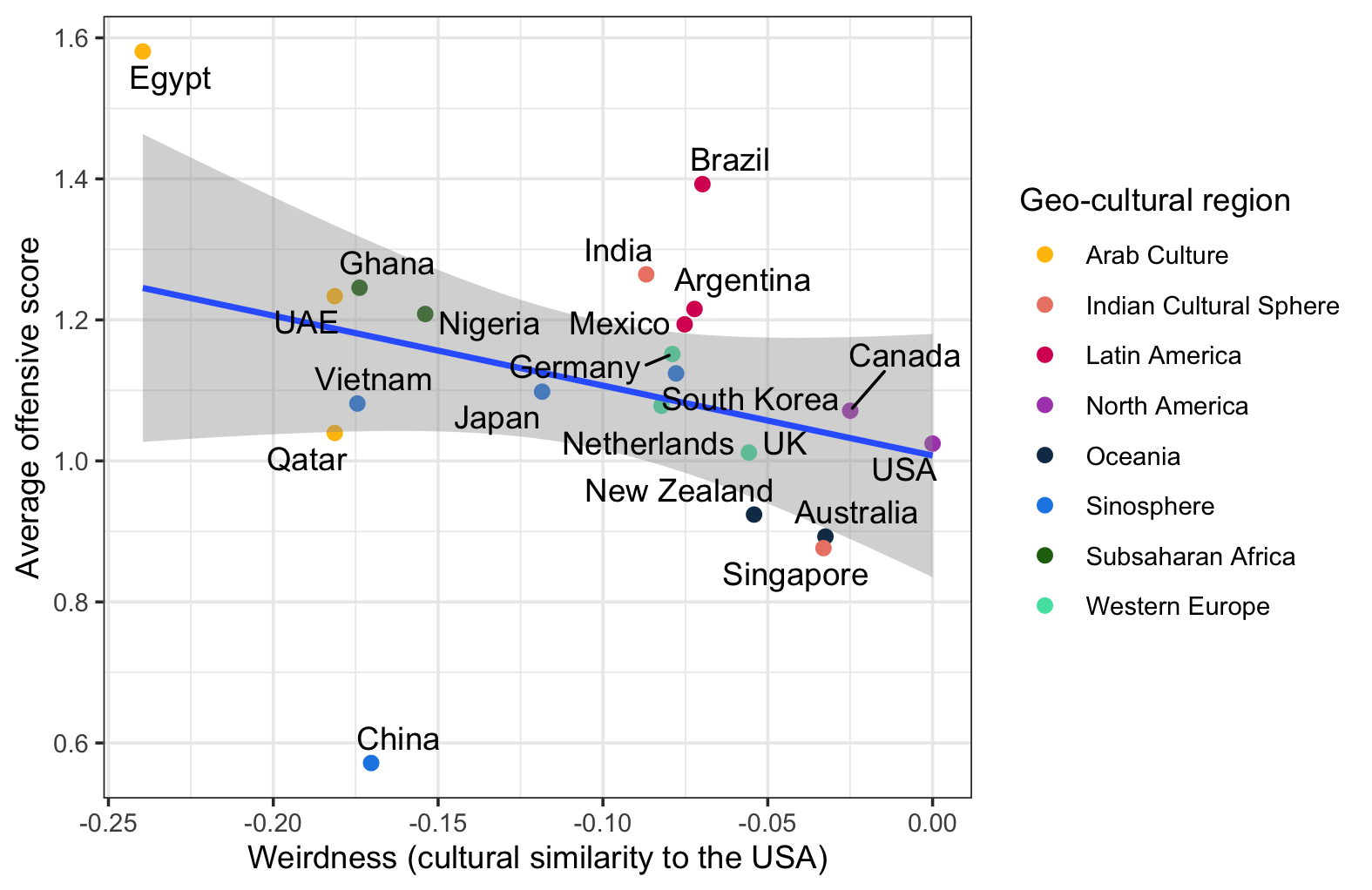}
    \caption{ 
    The association of cultural similarity to the USA with reported levels of offensiveness.
    }
    \label{fig:weird}
\end{figure}

To investigate whether the regional variances we observe can be explained by other demographic factors, we also performed the mixed-level regression analysis including annotators' age, gender, and self-reported socio-economic status, separately.
The effect of cross-regional differences holds even after accounting for annotators' age ($F$(7,7515) = 26.80, $p$ < .001), gender ($F$(7,7515) = 32.21, $p$ < .001), socio-economic status ($F$(7,7515) = 32.49, $p$ < .001) in the regression.

We further investigated whether providing a definition for offensiveness can significantly reduce the cross-regional differences. Cross-region variances are indeed lower when a definition is provided to participants ($F$(7, ) = 15.22 as compared to $F$(7, ) = 18.95). However, controlling for whether or not the definition is provided does not impact the observed results (a one-way ANOVA reported $F$(7,7515) = 31.54, $p$ < .001). In other words, even when annotators are asked to label based on a particular definition, there is still significant variance between participants of different regions.

While our regression analysis did account for country-level factors, we further analyzed how the trends we observe vary across countries within regions.
As Figure~\ref{fig:countries} shows, while participants from Egypt, Brazil, India, and Argentina are more likely to report offensive language (all four reporting an average offensiveness score above 1.2 across all items), participants from China labeled substantially fewer data points as offensive (with an average offensiveness score below 0.6). 
This stark difference in data collected from China could be, in part, due to country-specific norms on language use, or an artifact of that specific rater pool: e.g., more than two-third of the pool from China fall within the 30-50 years old range, while this age-group represents less than half the participants for all other countries (See Table~\ref{tab_stats}).\footnote{Because of this anomalous behaviour, we confirmed that our results hold even if we exclude data from China (which still retains over 350 participants from Sinosphere) in our region-level analysis.}

We also study how our country-level results correlate with established metrics that assess cultural differences between countries.
Participants from countries associated with WEIRD (Western, Educated, Industrialized, Rich, and Democratic) cultures \cite{henrich2010beyond} generally report fewer offensive labels (Figure \ref{fig:weird}). A one point increase in the WEIRDness score (collected from \cite{muthukrishna2020beyond} which measures cultural similarity to the United States) results in a 0.96 point decrease in the offensiveness score (although the effect is not significant with $p$ = 0.18).\footnote{Removing data from China from the analysis leads to significant impact of WEIRDness  on annotations ($p$ = 0.005), such that a one point increase in the WEIRDness score results in 1.54 decrease.}

We also study how our results correlate with established metrics of cultural differences documented at country-level along Hofstede's six cultural dimensions \cite{hofstede2011dimensionalizing}: \textit{Power Distance}, \textit{Uncertainty Avoidance}, \textit{Individualism/Collectivism}, \textit{Masculinity/Femininity}, \textit{Long/Short Term Orientation}, and \textit{Indulgence/Restraint}. While cultural tendency for seeking equality on power distribution or fulfilling one’s desires did not have a significant association with annotation behavior, we find that participants from countries that score higher on uncertainty avoidance ($\beta$ = 3.83, $p$ < .001), individualism ($\beta$ = 4.80, $p$ = .009), femininity ($\beta$ = 2.73, $p$ = .026), and short-term orientation ($\beta$ = 8.57, $p$ < .001) are more likely to report higher levels of offensiveness in their annotations. These results point to the importance of looking beyond the geographical regions, and instead at finer-grained values prevalent in specific countries.

%% file: sections/study2.tex
While Study 1 demonstrates significant cross-cultural differences in perceptions of offensiveness across regions and countries, cultural backgrounds of participants only partly explain the observed differences.
The mixed-effects model 
from Study 1
in fact shows that 39.5\% of the variance is due to individual differences while country and region difference explain only 2.9\% and 0.7\% of the variance, respectively.
Hence, we expand the scope of the analysis in Study 2, and 
include participant-level variables that have potential impact on annotation variances.

In particular, we consider the assessment of whether something is offensive or not as a matter of moral judgement, and hypothesize that participants' moral concerns play an important role in their assessments. To test this, we first examine whether participants' individual moral concerns that we measured through the Moral Foundation Questionnaire-2, mediate the effect of geo-cultural regions we observed on their perceptions of offensiveness in language.

\begin{figure}
    \centering
    
    \includegraphics[width=.235\textwidth]{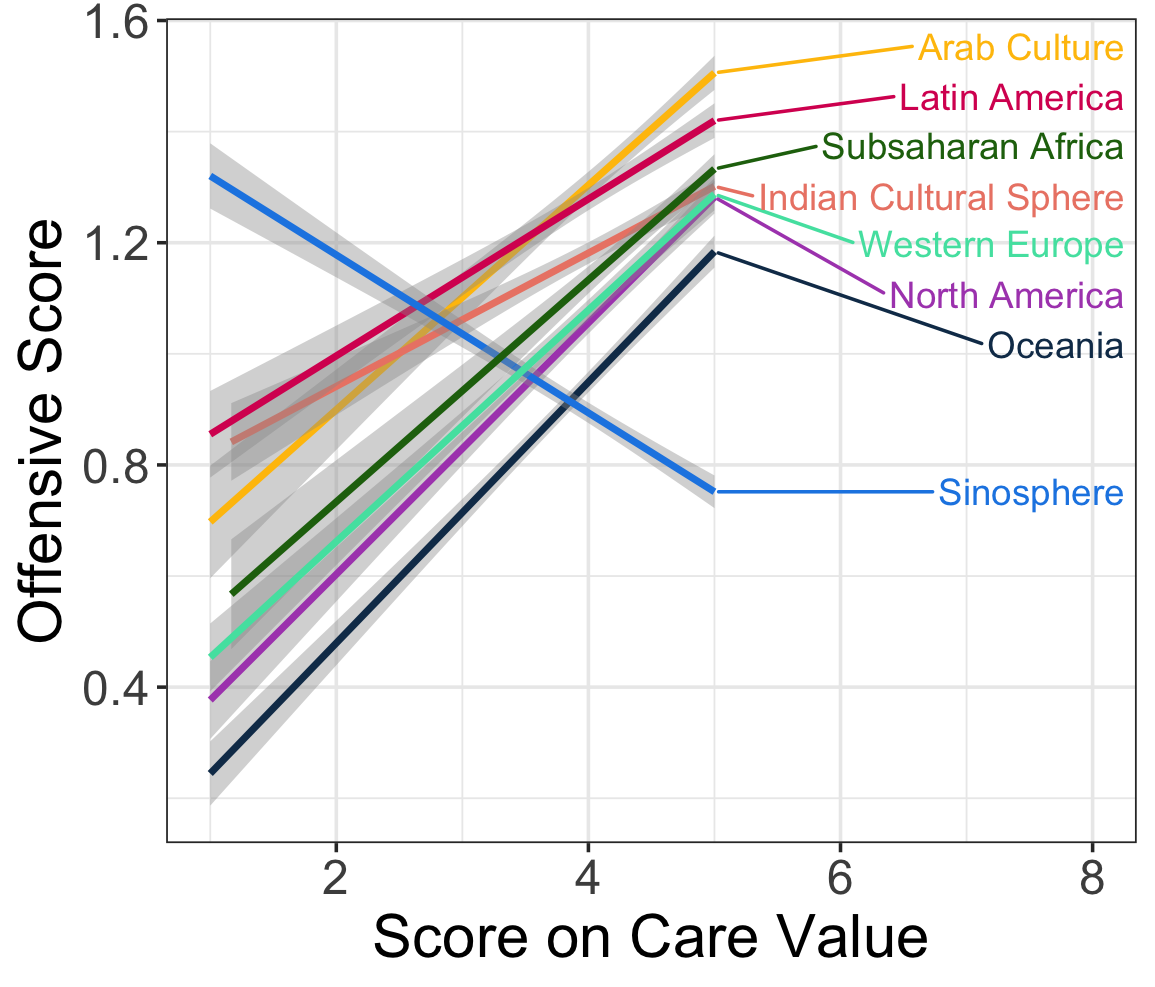}
    \includegraphics[width=.235\textwidth]{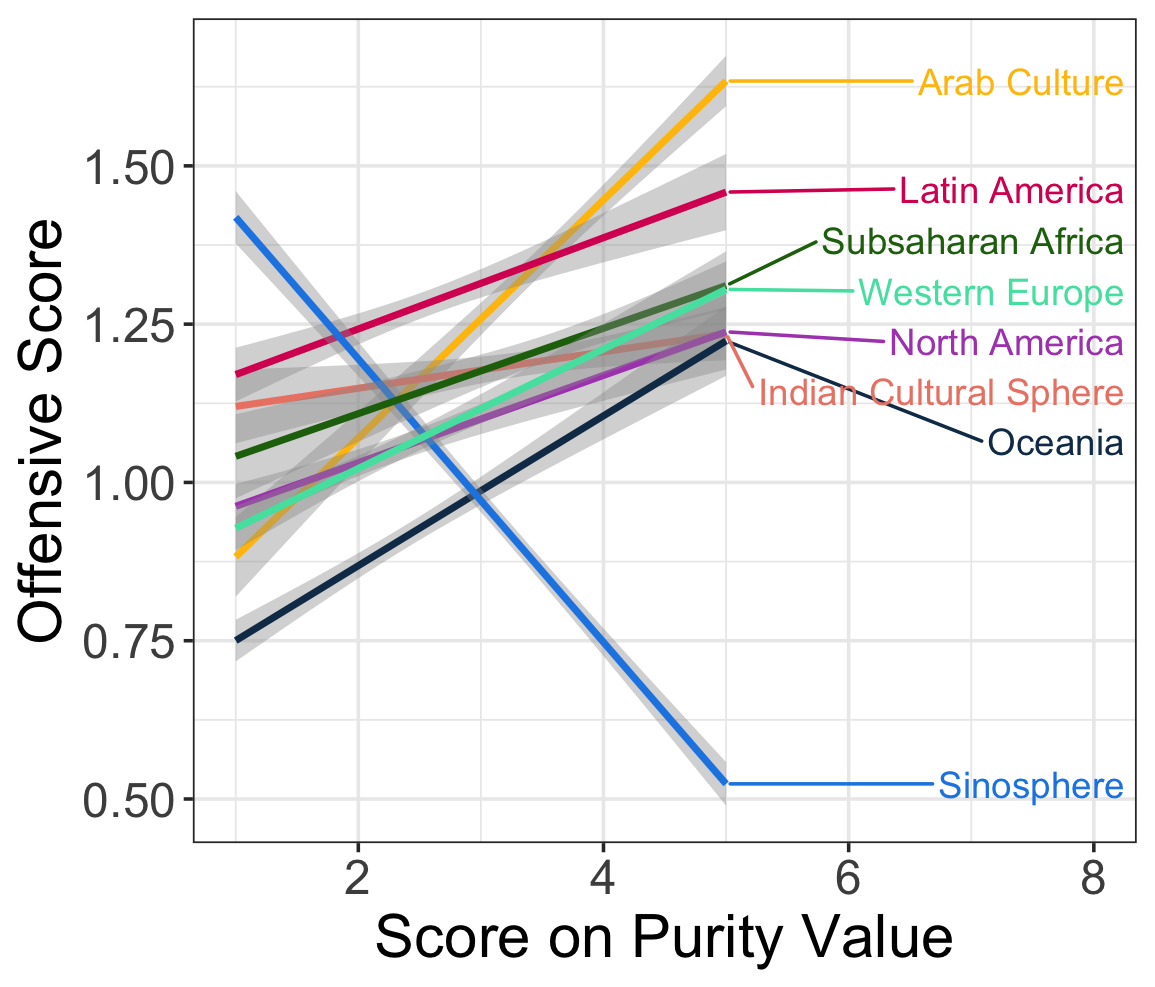}
    \caption{Annotators' scores on Care and Purity vary across regions and mediate the regional differences in annotating offensiveness. As the figures show one unit of increase in annotators' Care and Purity scores leads to varying levels of increase (or decrease in case of Sinosphere) in reported offensive scores.}
    \label{fig:mediation}
\end{figure}

We examine the mediation effect of each of the six moral foundations on the effects of regions on annotation differences. 
We consider each mediator separately since participants' score on each moral foundation is an independent score.
For each mediation test, we tested three associations:
(1) direct effect of independent variable (region) on dependent variable (offensiveness): confirmed in Study 1; (2) effect of independent variable (region) on the mediator (each moral foundation): six one-way ANOVA tests report significant effects of geo-cultural region on participants six morality scores (see Table \ref{tab:moral-region}); 
(3) combined effect of the independent variable and mediators on the dependent variable: when morality score variables are separately included to the association of geo-cultural region and annotations, the results show a significant mediating effect \cite{baron1986moderator} for Care ($ACME$\footnote{Average Causal Mediating Effect} = -0.034, $p$ < .001), and Purity ($ACME$ = -0.007, $p$ = .004).
In other words, annotators' perceptions of offensiveness that vary significantly across geo-cultural regions are significantly mediated by cultural differences in moral values regarding Care and Purity. As shown in Fig~\ref{fig:mediation}, an increase in annotators' Care and Purity scores leads to varying degrees of change in reported offensive scores in different regions; while Arab Culture demonstrates the highest positive change, Sinosphere demonstrates a chage in the opposite direction.\footnote{Note that we observe markedly different trends in Sinosphere,
which can be attributed to data from China noted before. Rerunning the mediation analysis excluding data from China shows that the effect of regions on annotation labels remains, with the mediation impact of Care (ACME= -0.043, p < .001), and Purity (ACME= -0.015, p < .001).}

We further assessed the impact of individual-level moral concerns that go beyond country-level concerns regarding Care and Purity on the annotations. To this end for each annotator $i$ from country $c$, we considered the moral scores (e.g., $Care_{i}$) as the summation of the average moral score of all annotators from their country (e.g., $Care_{c}$) and their deviation from the average (e.g., $Care_{i-c}$); such that: 

\begin{equation}
  Care_{i} = Care_{c} + Care_{i-c}  
\end{equation}

The results of two decomposition analyses for Care and Purity values show that in both cases the country-level moral concerns have less significant impact on annotations compared to the individual-level values. Specifically, in a mixed-level regression analysis with $Care_{i}$ and $Care_c$ as independent variables and annotation labels as the dependent variable, country-level Care score does not have a significant effect ($\beta$ = 0.27, $p$ = .087), while the individuals' deviation from their country has a significant impact on their annotations ($\beta$ = 0.14, $p$ < .001). In other words a 1-point increase in participants' Care score compared to their country's average is associated with 0.14 increase in the offensive score they assign to the items. A similar trend with smaller magnitude is observed for Purity scores where country-level Purity score does not have a significant effect ($\beta$ = 0.05, $p$ = .530), while the individuals' deviation from their country has a significant impact on their annotations ($\beta$ = 0.05, $p$ < .001). A 1-point increase in participants' Purity score compared to their country's average is associated with 0.05 increase in the offensive score they assign to the items. 

%% file: sections/study3.tex
Detecting objectionable content at scale is one of the core challenges in building responsible AI that effectively and equitably serves the global community. 
Our data collection relied on a geographically diverse pool of annotators and the results of Study 1 and 2 demonstrated how annotators from different geo-cultural regions with different moral values can provide varying perspectives about the offensiveness of language. However, most model training efforts are based on data annotation practices that
overlook these different perspectives about the task at hand.
To understand how this lack of representation impact data practices and models, we study the correlation of our collected labels with human-annotated labels provided in the original dataset (Jigsaw), and model predictions of a commercial tool for offensiveness detection (Perspective API). We ask two main questions: (1) whether labels provided in the dataset and trained model align more with annotators of specific regions, and (2) whether model labels are in higher agreement with annotators with particular moral concerns.

\begin{figure}
    \centering
    
    \includegraphics[width=.45\textwidth]{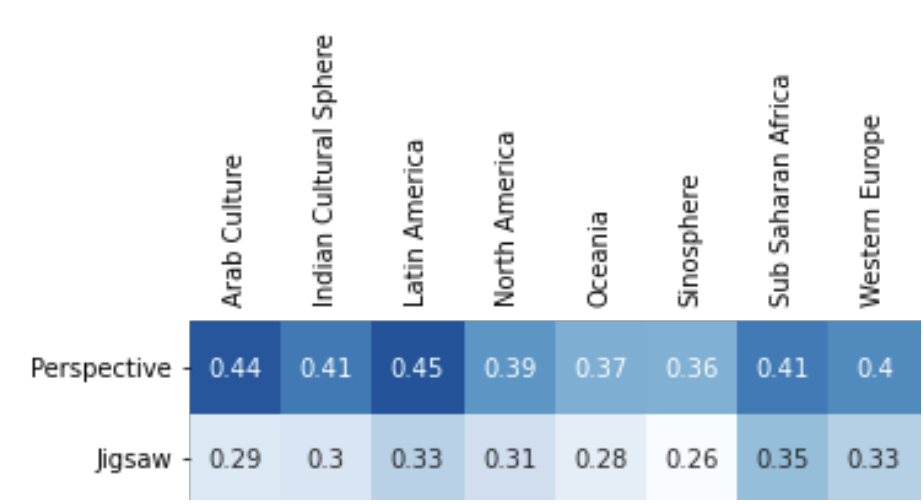}
    \caption{Correlation of labels provided in Perspective API with majority vote of annotators from different regions. Score interpretations can be summarized as: (0.00-0.20): neglectable correlation, (0.20-0.40): weak correlation, and (0.40-0.60): moderate correlation.}
    \label{fig:perspective-regions}
\end{figure}

Figure~\ref{fig:perspective-regions} presents the correlation of labels present in the dataset (Jigsaw) as well as those provided by the model (Perspective API), with the majority vote in our data collected from different regions. 
Across different geo-cultural regions, we observe a weak but significant correlation between labels in the Jigsaw dataset and majority labels we obtained from Sub Saharan Africa, Latin America, and Western Europe. On the other hand, we observe the least correlation with Sinosphere and Oceania.
It is important to note that the Jigsaw dataset is designed for the task of toxicity detection,\footnote{defined as ``a rude, disrespectful, or unreasonable comment that is likely to make one leave a discussion.''} in contrast to our data collection that centers around offensive language, as defined distinctively for half of our participants. Therefore, the relatively low correlation between our collected annotations and labels from Jigsaw should not come as a surprise. 

On the other hand, Perspective API labels have relatively higher correlation with majority votes of different regions although Perspective API, similar to Jigsaw's datasets, is trained to detect toxicity rather than offensiveness. The highest correlation is observed with annotations from Latin America and Arab Culture, and lowest correlation with Sinosphere and Oceania. Both the dataset and model has the least correlation with Sinosphere and Oceania, this can both be due to the fact that model and dataset in general report high levels of toxicity scores (Jigsaw $M$ = 0.43, $SD$ = 0.22, Perspective API $M$ = 0.39 , $SD$ = 0.26 on a 0 to 1 range), while annotators from Oceania and Sinosphere are the least likely to report high levels of offensiveness (Study 1). Moreover, higher correlation with Latin America, Arab Culture, and Sub Saharan Africa are potentially due to crowdworker recruitment approaches, which largely rely on workers from the Global South. This similarity is especially important since, we assume, Perspective API's model for English language is largely used by North American websites, therefore, the low correlation between model's predictions and the majority vote of North American participants may be an area of special concern.

Furthermore, to evaluate whether labels from Perspective API align with annotators with specific moral values, we calculated its level of agreement with each annotator. A regression model with this agreement score as the dependant variable and annotators' moral values as independent variables showed that Perspective API agrees more with annotators who score high on the Care value, such that a 1-point increase in the annotator's Care score leads to 0.027 ($SE$ = 0.004) increase in their agreement with Perspective API ($p$ < 0.001). 

In summary, these trends suggest not only that the perceptions of offensiveness are shaped by the socio-cultural backgrounds of the perceivers, but also that the datasets and models that have become standard in AI-based endeavors to detect and mitigate potentially offensive speech do preferentially reflect certain socio-cultural value systems over others.

%% file: sections/discussion.tex
Research on safety considerations of large language models has mostly focused on evaluations of model harms through crowdsourced benchmarks \cite{srivastava2022beyond,wang2021adversarial}. However, while annotators from different regions are shown to have different perspectives regarding this task \cite{salminen2018online}, current benchmarks fail to represent the cultural and individual variations in human moral judgements about generated language and model outputs. They also lack comprehensive understanding of human values and cultural norms that drive diversity of perspectives in annotations. 

Moreover, while the motivation behind safety evaluation tasks is to align language technologies with human values through a prescriptive approach \cite{ouyang2022training,kenton2021alignment}, the key role of culture in defining humans, and values is overlooked in benchmark creation and fine-tuning efforts. In fact, only a limited set of work in Machine Learning has investigated cultural differences in crowdsourcing pipelines for different tasks. Some examples include Chua et al. (2015), who investigated how cultural differences between requesters and contributors in creative crowdsourcing tasks influence both whether individuals decide to contribute as well as how requesters judge success \cite{chua2015impact}, and Joshi et al. (2016), who studied Indian linguists in a sarcasm annotation task, finding differences in their ability to detect sarcasm compared to ground truth datasets provided by American annotators \cite{joshi2016cultural}. 

Beyond cultural investigations, differences in annotation behavior have increasingly become a focus of study in machine learning. With the growth of sub-fields such as data-centric AI \cite{ng2021data,zha2023data}, researchers have put forth a range of explanations for disagreements in data annotation \cite{sandri2023don}, such as random variation as an artifact of human behavior \cite{krippendorff2008systematic}, as well as crowd worker and label schema quality \cite{aroyo2015truth,dumitrache2015crowdsourcing}.
Disagreements between specific social groups have also been studied, with scholars finding systematic differences in safety annotations rooted in gender \cite{cowan2003empathy}, race and sexuality \cite{goyal2022your}, community membership \cite{patton2019annotating}, and political views \cite{sap2021annotators}. As a result, researchers have problematized the notion of a universal gold standard for offensive or toxic language, and called for dataset and modeling approaches that preserve disagreements \cite{hovy2013learning,davani2022dealing,prabhakaran2021releasing}.

By conducting a cross-cultural experiment on identifying offensive language, in this paper we highlight two key factors that impact safety evaluations: (1) cross-cultural, and (2) psychological differences in detecting offensive language. We conducted a data collection effort with broad geographic coverage (21 countries from eight cultural regions), having each item labeled for offensiveness by at least three annotators from each region. Our multi-level analysis of responses provides important insights into how geo-cultural factors and individual moral factors influence perceptions of offense in language:

\subsection*{Geo-Cultural Factors}
The results of Study 1 provide strong evidence for cross regional differences in annotating offensiveness in language. Countries associated with Arab Culture, Latin America, Sub-Saharan Africa, and Indian Cultural Sphere, were more likely to annotate items as offensive compared to the other four regions, i.e., North America, Western Europe, Oceania, and Sinosphere; with individuals from Sinosphere, and specifically China, providing the least offensive labels. Often referred to as the Global South, the four regions with higher reports of offensive, have been compared to western countries in terms of their wide cultural differences. 
Considering the WEIRDness of countries (their degree of similarity from the USA culture), while not having a significant impact on provided labels, is associated with lower offensiveness labels. In other words participants from countries which are more culturally similar to the USA (this includes North America, Western European Countries, Oceania, and some countries across other regions such as Singapore) are less sensitive to offensive content.

We also investigated the impact of the six cultural factors introduced by Hofstede, on countries responses to offensive content. While cultural tendency for seeking equality on power distribution (Power-Distance Index) and cultural tendency for fulfilling one's desires (Indulgence) did not have significant association with annotation behaviors, cultures with a higher tendency to avoid uncertainty and ambiguity (Uncertainty Avoidance), more gender role differentiation (Femininity vs. Masculinity), and stronger emphasis on the present than the future (short-term orientation) are more sensitive to offensive language. 
These results provide finer-grained insights into values that may shape individuals' tolerance for offensive language that goes beyond regional groupings. 

\subsection*{Moral Factors}
We next investigated the mediating impact of moral concerns on the regional differences in perceiving offensiveness. We expected individual and cultural differences in moral values (measured through the moral foundations questionnaire) to explain the observed regional and cultural differences. Our results showed that regions with higher moral values regarding Caring for others and Purity are more sensitive to offensive language. Care as a main definitive foundation discussed in different moral theories, includes ``intuitions about avoiding emotional and physical damage to another individual'' \cite{atari2022morality}.  Care values have lower variance across regions compared to the other foundations (Table.~\ref{tab:moral-region}), with most people scoring high on this foundation, a high score on Care helps explain higher offensiveness labels across the regions. On the other hand, the Purity foundation, with higher variance across regions, also has a positive association with high offensiveness labels. Purity refers to ``intuitions about avoiding bodily and spiritual contamination and degradation'' \cite{atari2022morality}.


As we demonstrate, determining the offensiveness of language is 
a moral judgment shaped by cultural factors.
However, research on offensive language detection typically disregards sociodemograhic and cultural variations in data annotation.
Moreover, the need to assess the cultural values and moral norms takes on even more weight in the context of large language models, which enable a scale of language production beyond that of the human-generated content typically analyzed in social media contexts. These large language models rely on fine-tuning processes that involve human-generated annotations and feedback that are inherently shaped by cultural and moral views. As a result, generative models may perform in ways that differentially align with the norms of different global populations. 
Indeed, Santy et al. (2023) \cite{santy2023nlpositionality} analyzed dataset ground truth labels and model outputs for a social acceptability and hate detection task, finding greater alignment with annotations provided by Western, White, college-educated, and younger individuals. 

Our findings confirm that disagreements in annotating offensiveness, with key applications in evaluating AI safety, have cultural, and psychological roots often disregarded in current efforts. 
These findings call for culturally-informed data collection and model evaluation  effort, ensuring that they reflect the values and norms in the communities affected by model deployments. It is, therefore, essential to diversify rater pools of data annotator and model evaluators to incorporate various perspectives of language to enhance modeling processes responsibly.

Moreover, inclusion of diverse opinions goes beyond data collection and expansion of rater pools; we believe that cultural considerations play a pivotal role in defining AI-related harms and devising more effective safety protocols and paradigms.
Nevertheless, it is crucial to recognize that defining AI harms according to cultural values can conflict with social equity for marginalized groups within a given society. As discussed by \cite{hershcovich-etal-2022-challenges}, our findings supports NLP  efforts that simultaneously uphold cultural values while actively mitigating cultural biases. This is only possible with active community engagement for aligning AI models with cultural norms, along with relying on international regulatory efforts to safeguard human rights of vulnerable individuals within communities \cite{prabhakaran2022human}.

Lastly, our findings contributes a vital perspective to ongoing discourse regarding the necessity for AI models to align with human values. Considering the diverse values and perspectives present regions, cultures, and individuals, the critical question to ask is whose values should the models align with. Instead of mainly focusing on demographics, our findings suggests that aligning AI models with moral foundations that individuals and groups are concerned about potentially provides an effective and efficient approach towards value alignment efforts for the pluralistic world we live in. 


%% file: sections/appendix.tex
\input{img/demo-tab}

\paragraph{Participants}
Our data collection process aimed to recruit a balanced set of annotators from eight geo-cultural regions, namely, Northern America, Latin America, Arab Cultures, Sinosphere, Oceania, Sub Saharan Africa, Indian Cultural Sphere, and Western Europe.\footnote{Our efforts for collecting data from Eastern Europe halted because of the Russia's war in Ukraine.} We recruited at least 450 participants from each region ($N_{participants}$ = 4309), 
represented by 2-4 countries, 
with at least 100 participants per country,
except for South Korea and Qatar where 
we managed to recruit only a smaller number of raters. See Table~\ref{tab_stats} for country-level statistics.
While typical raters recruited through a dedicated annotation platform or service possess specialized training or experience that may influence how they complete tasks, most platforms feature a relatively limited degree of cultural representation within their pools. As a result, they are less appropriate for specifically and robustly evaluating cultural differences and their relationship to annotation judgments.


\paragraph{Annotation}
Each participant was assigned to annotate offensiveness of 40 textual items with a value between 0 (not offensive at all) and 5 (extremely offensive). We asked participants to decide whether they perceived them as offensive, noting that ``whether something is offensive or not may depend on the context in which this message is exchanged, which we are not providing you in these questions. Please make the best assessment based on the information/context you have from the provided text.'' Moreover, half of the participants were provided with a note that defined \textit{extremely offensive language} as ``profanity, strongly impolite, rude or vulgar language expressed with fighting or hurtful words in order to insult a targeted individual or group.'' Other participants were expected to label items based on their own definition of offensiveness. Participants’ common understanding of offensiveness was tested by 5 (non-offensive) controlled questions randomly distributed among the 40-items annotation process. Participants with at least one wrong answer were excluded.

\paragraph{Stimuli}
Textual items were selected from Jigsaw's Toxic Comments Classification dataset \cite{Jigsaw-toxic}, and the Unintended Bias in Toxicity Classification dataset \cite{Jigsaw-bias}.
We set our data collection goal to increase the chance of evoking disagreements, and involving different moral judgments in annotations. To address the former, we randomly selected items from Jigsaw's Toxic Comments Classification dataset, prioritizing those with high disagreement levels. To measure disagreements on each item, we averaged annotators' toxicity scores ranging from 0 (lowest toxicity) to 1 (highest toxicity). We chose items with a normal distribution centered around a toxicity score of 0.5 (indicating highest disagreement) with a standard deviation of 0.2.
To select items that would potentially evoke moral reasoning we collected items from Jigsaw's Unintended Bias in Toxicity Classification dataset with moral sentiment, labeled by on a supervised moral foundation tagger trained on the Moral Foundations Twitter Corpus \cite{hoover2020moral}. We also collected a third set of items each mentioning at least one social group (this information is provided in the Jigsaw's raw data), to assess annotation of language that is potentially offensive toward different social groups.

\begin{table}[ht]
    \centering
    \scalebox{.9}{
    \begin{tabular}{m{4em} m{10em} m{4em}}
    \toprule
        Foundation && $F$(7, 4287) \\\toprule
        Care & \includegraphics[width=.25\textwidth]{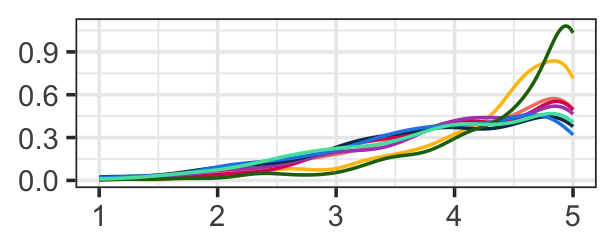} & 34.48*  \\\hline
        Equality & \includegraphics[width=.25\textwidth]{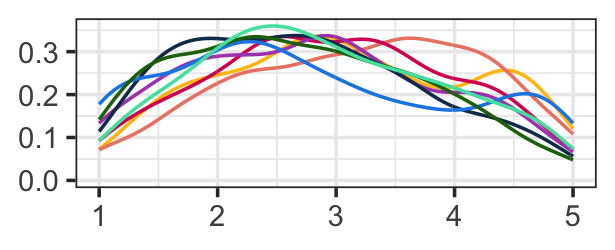} & 13.37* \\\hline
        Propor. & \includegraphics[width=.25\textwidth]{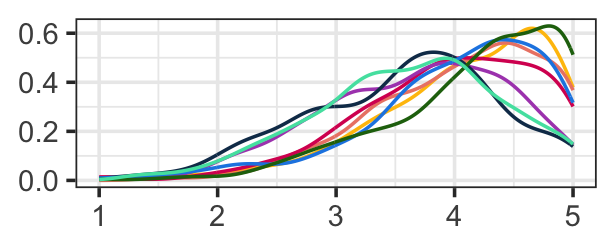} & 51.24* \\\hline
        Authority & \includegraphics[width=.25\textwidth]{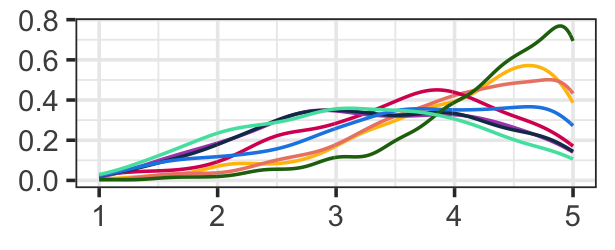} & 102.25* \\\hline
        Loyalty & \includegraphics[width=.25\textwidth]{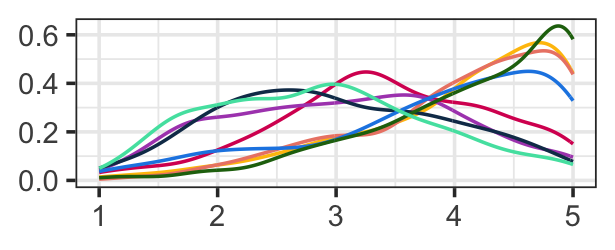} & 158.30* \\\hline
        Purity & \includegraphics[width=.25\textwidth]{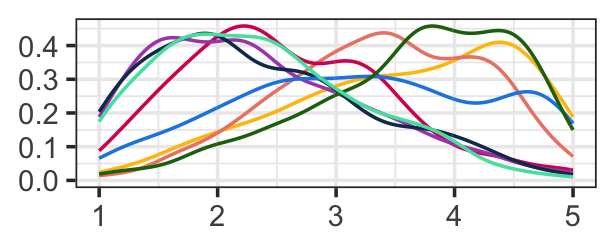} & 203.55* \\\hline
        \includegraphics[width=.45\textwidth]{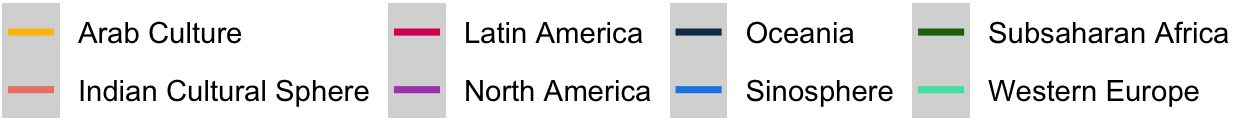}&  &  \\\hline
    \end{tabular}}
    \caption{Distribution of the moral foundations scores and the results of one-way ANOVA analysis conducted for each moral foundation across regions. *means the $p$-value of the analysis is lower than .001}
    \label{tab:moral-region}
\end{table}

\paragraph{Regions and Countries}
Our selected list of geo-cultural regions and countries within regions is not meant to be exhaustive, rather just to make sure that our study is done on a set of countries with diverse cultural histories. Each region listed has countries and sub-regions that have distinct cultural practices, and it is wrong to assume that the country we choose would comprehensively represent that region. Similarly, the countries listed are meant as likely places to collect data from, based on familiarity with previous data collection efforts, which potentially reflect the power structures existing within those regions. Also, each country is rarely a monolith in terms of culture (e.g., India has diverse subcultures, Australia being characterized as a ``Western'' culture erases the vibrant Australian Aboriginal culture). Data collected would also reflect the local disparities in who tends to be in the social strata that maximally overlaps with the data-collection workforce in those respective regions, and what subcultures they represent.


\paragraph{Moral Foundation Questionnaire-2}
After annotating items, participants were asked to fill out the Moral Foundation Questionnaire \citep{atari2022morality}, which includes 36 statements, each concerning one moral value by extending the the moral foundation theory \cite{graham2009liberals} (i.e., \textit{Care}, \textit{Equality}, \textit{Proportionality}, \textit{Loyalty}, \textit{Authority}, and \textit{Purity}). Participants' concerns regarding each item were captured on a 5-point Likert scale, and items where aggregated to calculate 6 moral scores for each participant.

\paragraph{Test Items}
We used participants' response to 5 test items to evaluate their understanding of offensiveness. These items (listed below) were unanimously labeled as \textit{not offensive at all} by authors and were used as an attention check to remove participants who annotated them otherwise. 

\begin{itemize}[noitemsep,nolistsep]
    \item Could you explain your point in a little more detail please.
    \item The NY Times yesterday had a rebuttal by the Iranian UN ambassador.
    \item Ditto to everything said here.
    \item Just looked at the link. Impressive and missed opportunity.
    \item Don't be so hard on yourself. Your life will go on.
\end{itemize}

\paragraph{Data Cleaning}
We selected thresholds for the amount of time needed to finish the survey and removed annotators who performed the task either quicker or slower than the expectation. Annotators with similar answers to all items were also removed from the data.

\paragraph{Mediation}
We perform the mediation analysis on the annotator level by aggregating annotation labels provided from each annotator. While annotators labeled different parts of the dataset (each annotating 35 out of the 5k items) equal number of annotators from each region were assigned to label the same set. In other words for each set of 35 items, there are 3 annotators from each region that labeled the whole set.

%% file: img/demo-tab.tex
\begin{table*}[th!]
\centering
\scalebox{.9}{
\begin{tabular}{p{2.2cm}p{2cm}ccccccccc}\toprule
&&& \multicolumn{4}{c}{\textbf{Gender}} && \multicolumn{3}{c}{\textbf{Age}}\\\cline{4-7}\cline{9-11}
\textbf{Region}                         & \textbf{Country}         & \multicolumn{1}{l}{\textbf{\#}} & \multicolumn{1}{l}{\textbf{Man}} & \multicolumn{1}{l}{\textbf{Woman}} & \multicolumn{1}{l}{\textbf{NB}} & \multicolumn{1}{l}{\textbf{PNS}} && \multicolumn{1}{l}{\textbf{18 -- 30}} & \multicolumn{1}{l}{\textbf{30 -- 50}} & \multicolumn{1}{l}{\textbf{50+}}
 \\\midrule
\multirow{3}{*}{Arab Culture}           & Egypt                    & 225                                              & 61.80\%                          & 36.40\%                            & 0.40\%                                  & 1.30\%                                         && 55.60\%                              & 20.90\%                              & 23.60\%                           \\
                                        & Qatar                    & 57                                               & 64.90\%                          & 33.30\%                            & 0.00\%                                  & 1.80\%                                         && 63.20\%                              & 31.60\%                              & 5.30\%                            \\
                                        & UAE     & 234                                              & 55.60\%                          & 44.40\%                            & 0.00\%                                  & 0.00\%                                         && 46.20\%                              & 44.00\%                              & 9.80\%                            \\\hline
\multirow{2}{*}{\shortstack[l]{Indian Cultural \\ Sphere}} & India                    & 444                                              & 57.00\%                          & 43.00\%                            & 0.00\%                                  & 0.00\%                                         && 46.60\%                              & 34.20\%                              & 19.10\%                           \\
                                        & Singapore                & 110                                              & 50.00\%                          & 49.10\%                            & 0.90\%                                  & 0.00\%                                         && 27.30\%                              & 41.80\%                              & 30.90\%                           \\\hline
\multirow{3}{*}{\shortstack[l]{Latin\\America}}          & Argentina                & 149                                              & 50.30\%                          & 47.70\%                            & 2.00\%                                  & 0.00\%                                         && 52.30\%                              & 34.90\%                              & 12.80\%                           \\
                                        & Brazil                   & 237                                              & 47.30\%                          & 52.70\%                            & 0.00\%                                  & 0.00\%                                         && 57.40\%                              & 27.00\%                              & 15.60\%                           \\
                                        & Mexico                   & 163                                              & 51.50\%                          & 48.50\%                            & 0.00\%                                  & 0.00\%                                         && 54.00\%                              & 36.80\%                              & 9.20\%                            \\\hline
\multirow{2}{*}{\shortstack[l]{North\\America}}          & Canada                   & 378                                              & 37.60\%                          & 61.90\%                            & 0.50\%                                  & 0.00\%                                         && 41.00\%                              & 36.80\%                              & 22.20\%                           \\
                                        & USA & 173                                              & 45.10\%                          & 52.60\%                            & 1.20\%                                  & 1.20\%                                         && 62.40\%                              & 20.80\%                              & 16.80\%                           \\\hline
\multirow{2}{*}{Oceania}                & Australia                & 184                                              & 39.70\%                          & 58.70\%                            & 1.60\%                                  & 0.00\%                                         && 25.50\%                              & 50.00\%                              & 24.50\%                           \\
                                        & New Zealand              & 333                                              & 39.00\%                          & 59.80\%                            & 1.20\%                                  & 0.00\%                                         && 34.20\%                              & 38.70\%                              & 27.00\%                           \\\hline
\multirow{4}{*}{Sinosphere}             & China                    & 176                                              & 37.50\%                          & 62.50\%                            & 0.00\%                                  & 0.00\%                                         && 14.20\%                              & 66.50\%                              & 19.30\%                           \\
                                        & Japan                    & 100                                              & 70.00\%                          & 30.00\%                            & 0.00\%                                  & 0.00\%                                         && 13.00\%                              & 38.00\%                              & 49.00\%                           \\
                                        & South Korea              & 43                                               & 58.10\%                          & 41.90\%                            & 0.00\%                                  & 0.00\%                                         && 27.90\%                              & 48.80\%                              & 23.30\%                           \\
                                        & Vietnam                  & 221                                              & 53.80\%                          & 41.20\%                            & 5.00\%                                  & 0.00\%                                         && 71.50\%                              & 23.50\%                              & 5.00\%                            \\\hline
\multirow{2}{*}{\shortstack[l]{Sub-Saharan\\Africa}}     & Ghana                    & 164                                              & 67.10\%                          & 32.90\%                            & 0.00\%                                  & 0.00\%                                         && 74.40\%                              & 22.00\%                              & 3.70\%                            \\
                                        & Nigeria                  & 366                                              & 54.40\%                          & 45.10\%                            & 0.30\%                                  & 0.30\%                                         && 54.10\%                              & 33.10\%                              & 12.80\%                           \\\hline
\multirow{3}{*}{\shortstack[l]{Western\\Europe}}         & Germany                  & 109                                              & 52.30\%                          & 46.80\%                            & 0.90\%                                  & 0.00\%                                         && 51.40\%                              & 24.80\%                              & 23.90\%                           \\
                                        & Netherlands              & 138                                              & 52.20\%                          & 45.70\%                            & 2.20\%                                  & 0.00\%                                         && 61.60\%                              & 21.00\%                              & 17.40\%                           \\
                                        & United Kingdom           & 305                                              & 40.30\%                          & 59.00\%                            & 0.70\%                                  & 0.00\%                                         && 38.70\%                              & 38.00\%                              & 23.30\%   \\\bottomrule                       
\end{tabular}}
\caption{Socio-demographic distribution of participants in our study in different regions and countries. (PNS = Prefer not to say, and NB = Non-binary)}

\label{tab_stats}
\end{table*}